\documentstyle[preprint,revtex]{aps}
\begin{document}
\draft
\begin{title}
Riemann's theorem for quantum tilted rotors
\end{title}
\author{G.\ Rosensteel and A.L.\  Goodman}
\begin{instit}
Department of Physics, Tulane University, New Orleans, Louisiana
70118
\end{instit}
\begin{abstract}
The angular momentum, angular velocity, Kelvin circulation, and
vortex velocity vectors of a quantum Riemann rotor are proven to be
either (1) aligned with a principal axis or (2) lie in a principal plane of
the inertia ellipsoid.  In the second case, the ratios of the components
of the Kelvin circulation to the corresponding components of the
angular momentum, and the ratios of the components of the angular
velocity to those of the vortex velocity are analytic functions of the
axes lengths.
\end{abstract}

\narrowtext

A classical Riemann rotor is a uniform density fluid with an ellipsoidal
boundary and a velocity field that is a linear function of position.
Riemann fluids model rotating stars and galaxies,\cite{CHA69,LEB67}
spinning gas clouds,\cite{DYS68} and rotating
nuclei.\cite{CUS68,ROS88}  Since linear velocity fields span the
dynamical continuum from rigid rotation to irrotational flow, the
Riemann model is sufficiently general to model most collective
rotational systems.  The vector observables that measure the
character of the rotation are the angular momentum $\vec{L}$ and the
Kelvin circulation $\vec{C}$;\cite{ROS88,WEA76,GUL76,ROS76} the
vector variables conjugate to $\vec{L}$ and $\vec{C}$ are the angular
velocity $\vec{\omega}$ and the vortex velocity $\vec{\lambda}$.
Tilted nuclear rotors for which $\vec{\omega}$ is not aligned with a
principal axis are a topic of continuing
interest.\cite{FRI87,GOO92,NAZ92,THO62,MAR79,MIK78,VAS80,KER81,CUY87,KLE
79,DON90,FRA91}

Recently, a quantum theory of Riemann rotors was formulated by
simultaneous angular and vortex cranking of the nuclear mean field
Hamiltonian,
\begin{equation}
H_{\vec{\omega} \vec{\lambda}}  = H_{0} - \vec{\omega}\cdot \vec{L} +
\vec{\lambda}\cdot \vec{C},
\end{equation}
where the mean field is approximated by the anistropic oscillator
potential,
\begin{equation}
H_{0} = - \frac{\hbar^{2}}{2m} \bigtriangleup +
\frac{1}{2}m(\omega^{2}_{x} x^{2}+\omega^{2}_{y}
y^{2}+\omega^{2}_{z} z^{2}).
\end{equation}
Inglis's cranking formula determines the collective energy of an
\mbox{$A$-nucleon} system,\cite{ING54,RIN80}
\begin{equation}
T(\vec{\omega},\vec{\lambda}) = \sum_{ph}\frac{\mid \langle p \mid
\vec{\omega}\cdot \vec{L} - \vec{\lambda}\cdot \vec{C}\mid h \rangle
\mid^{2}}{\epsilon_{p}-\epsilon_{h}}. \label{TINGLIS}
\end{equation}
Self-consistency of the mean field with the density distribution requires
that
\begin{equation}
\omega_{x}N_{x} = \omega_{y}N_{y} = \omega_{z}N_{z}, \label{SELF}
\end{equation}
where $N_{k} = \sum (n_{k}+1/2)$ denotes the total number of quanta
in the k$^{th}$ direction.  The following semiclassical correspondence
theorem has been established:\cite{ROS92} At self-consistency, the
``Inglis'' collective energy, Eq.\ (\ref{TINGLIS}), equals the classical
value for the kinetic energy of a Riemann rotor,\cite{CHA69}
\FL
\begin{equation}
T(\vec{\omega},\vec{\lambda}) = \frac{I_{0}}{4} \sum_{k=1}^{3}
[(a_{i}^{2}+a_{j}^{2})(\omega_{k}^{2}+\lambda_{k}^{2}) -
4a_{i}a_{j}\omega_{k}\lambda_{k}],
\end{equation}
where $i,j,k$ are cyclic, $a_{1},a_{2},a_{3}$ are the axes lengths of
the inertia ellipsoid in units of R, and the moment of inertia of a sphere
of radius R and total mass M=mA is $I_{0}=(2/5)MR^{2}$.  If the
nuclear volume is set equal to the spherical value $4\pi R^{3}/3$, then
the product of the dimensionless axes lengths equals unity,
$a_{1}a_{2}a_{3}=1$.  Note that if the vortex velocity vanishes, then
the self-consistent collective energy equals the rigid body energy, a
\mbox{well-known} result.\cite{BOH75}  In addition, the quantum
expectations of the angular momentum and Kelvin circulation equal
their classical Riemann rotor values,\cite{CHA69}
\begin{eqnarray}
\langle L_{k}\rangle & = & \frac{I_{0}}{2} \left[
(a_{i}^{2}+a_{j}^{2})\omega_{k} - 2a_{i}a_{j}\lambda_{k}\right]
\nonumber \\
\langle C_{k}\rangle & = & \frac{I_{0}}{2} \left[ 2a_{i}a_{j}\omega_{k} -
(a_{i}^{2}+a_{j}^{2})\lambda_{k}\right] \label{expLK}.
\end{eqnarray}
These expectations are given by derivatives of the kinetic energy with
respect to the angular velocity and the vortex velocity\cite{FEY39}
\begin{equation}
\langle L_{k}\rangle = \left( \frac{\partial T}{\partial \omega_{k}}\right),
\mbox{ } \langle C_{k}\rangle = -\left( \frac{\partial T}{\partial
\lambda_{k}}\right). \label{LKDT}
\end{equation}
The collective energy may be expressed as
\begin{equation}
T(\vec{\omega},\vec{\lambda}) = \frac{1}{2}\left( \vec{\omega}\cdot
\langle \vec{L} \rangle - \vec{\lambda}\cdot \langle \vec{C}\rangle
\right) .
\end{equation}

The energy in the rotating intrinsic frame for ordinary ``Inglis'' cranking
of the angular velocity vector is minimized with respect to orientation
when the angular velocity and angular momentum vectors are
parallel.\cite{NAZ92,THO62,FRA91}  Because the vortex velocity is
independent of the angular velocity, minimization of the intrinsic
energy $\tilde{E}(\vec{\omega},\vec{\lambda})=\langle
H_{\vec{\omega} \vec{\lambda}}\rangle$ with respect to the orientation
of $\vec{\lambda}$ also requires that the vortex velocity and the Kelvin
circulation vectors are parallel,
\begin{mathletters}
\begin{eqnarray}
\vec{\omega}\times \langle \vec{L} \rangle & = & 0 \label{wtimesL} \\
\vec{\lambda}\times \langle \vec{C} \rangle & = & 0 . \label{PARALLEL}
\end{eqnarray}
\end{mathletters}
This is proven by computing the change in the intrinsic energy when
the orientation of the vortex velocity is shifted infinitesimally from
$\vec{\lambda}$ to $\vec{\lambda}+\delta \vec{\lambda} =
\vec{\lambda} + \epsilon \,\hat{n}\times \vec{\lambda}$, where
$\hat{n}$ is an arbitrary unit vector,
\FL
\begin{eqnarray}
\tilde{E}(\vec{\omega},\vec{\lambda}+\delta \vec{\lambda}) -
\tilde{E}(\vec{\omega},\vec{\lambda}) & = & \epsilon (\hat{n}\times
\vec{\lambda})\cdot \langle \vec{C}\rangle + O(\epsilon^{2})\\
& = & \epsilon \,\hat{n}\cdot (\vec{\lambda}\times \langle \vec{C}\rangle
)+ O(\epsilon^{2})
\end{eqnarray}
Hence, at equilibrium, $\hat{n}\cdot (\vec{\lambda}\times \langle
\vec{C}\rangle ) = 0$ for all directions $\hat{n}$, or
$\vec{\lambda}\times \langle \vec{C}\rangle = 0$.

Riemann\cite{RIE60} proved in 1860 that the classical rotors in
equilibrium fall into three classes:
\begin{enumerate}
\item Rigid rotors ($\lambda$=0) which encompasses the Maclaurin
spheroids and the Jacobi triaxial ellipsoids,
\item S-type ellipsoids for which the directions of $\vec{\omega}$ and
$\vec{\lambda}$ are aligned with a principal axis, and
\item Tilted ellipsoids for which the directions of $\vec{\omega}$ and
$\vec{\lambda}$ lie in a principal plane.
\end{enumerate}
The case where $\vec{\omega}$ and/or $\vec{\lambda}$ do not lie in a
principal plane is specifically excluded.  It is remarkable that
Riemann's theorem is also true for quantum cranked Riemann rotors.
To prove the theorem, substitute the explicit formulae for the
expectations of the angular momentum and Kelvin circulation, Eq.\
(\ref{expLK}), into the parallelism conditions, Eqs.\
(\ref{wtimesL},\ref{PARALLEL}).  If the angular velocity vector
$\vec{\omega}$ is neither aligned with a principal axis nor lies in a
principal plane, then $\omega_{1}$, $\omega_{2}$, $\omega_{3}$ are
all nonzero and the parallelism conditions constitute a set of six
simultaneous equations in the three unknown ratios,
$\lambda_{1}/\omega_{1}$, $\lambda_{2}/\omega_{2}$,
$\lambda_{3}/\omega_{3}$.  It can be shown that only four of these
equations are independent.  Therefore, this simultaneous system is
overdetermined for the three unknown ratios.  Since the assumption
that all three components $\omega_{k}$ are nonzero implies a
contradiction, one of the angular velocity components must vanish,
say $\omega_{1}$.  But, writing out the \mbox{$y$-component} of Eq.\
(\ref{wtimesL}), $0=\omega_{3}\langle L_{1} \rangle = -
2a_{2}a_{3}\omega_{3}\lambda_{1}$, one concludes that
$\lambda_{1}$ must also vanish.  Hence, $\vec{\omega}$ and
$\vec{\lambda}$ lie in a principal plane of the inertia ellipsoid.  If
$\omega_{2}$ and $\omega_{3}$ are both nonzero, then the
parallelism conditions produce two independent equations in two
unknown ratios whose solution is
\begin{eqnarray}
\lambda_{2}/\omega_{2} & = & (4a_{1}^{2}-a_{2}^{2}+a_{3}^{2}\pm
q)/4a_{1}a_{3} \nonumber \\
\lambda_{3}/\omega_{3} & = & (4a_{1}^{2}+a_{2}^{2}-a_{3}^{2}\pm
q)/4a_{1}a_{2},
\end{eqnarray}
where
\begin{equation}
q^{2} = (4a_{1}^{2}-(a_{3}+a_{2})^{2})(4a_{1}^{2}-(a_{3}-a_{2})^{2}).
\end{equation}
If only one of the components of $\vec{\omega}$ is nonzero, then the
angular velocity, vortex velocity, angular momentum, and Kelvin
circulation vectors are all aligned with a single principal axis.  The
ratio $\lambda /\omega$ is undetermined for such an S-type ellipsoid.
This completes the proof of Riemann's theorem.

Since $q$ cannot be imaginary, there are only three types of tilted
rotors.  Choosing an ordering of the axes lengths in the principal
plane, say $a_{3}\geq a_{2}$, yields the following:
\begin{itemize}
\item {\em Type I}. $2a_{1}\geq a_{3}+a_{2}$
\item {\em Type II}. $2a_{1}\leq a_{3}-a_{2}$ and $a_{2}\leq a_{1}$
\item {\em Type III}. $2a_{1}\leq a_{3}-a_{2}$ and $a_{1}\leq a_{2}$.
\end{itemize}
Type II and Type III tilted rotors are ultradeformed \mbox{prolate-like}
solutions for which the ratio of the longest to the shortest axis is at
least three to one.  Type I solutions are tilted \mbox{oblate-like} rotors;
these have been studied in the classical macroscopic approximation
where the potential energy is a sum of the attractive surface energy
plus the repulsive Coulomb potential.\cite{ROS88}

\end{document}